\documentclass[a4paper,11pt]{article}
\usepackage{jinstpub} 

\title{Status of a point-like neutron generator development}

\author[a]{S.V.~Golubev,}
\author[a,b]{V.A.~Skalyga,}
\author[a,b,1]{I.V.~Izotov,\note{Corresponding author.}}
\author[a]{S.V.~Razin,}
\author[a]{R.A.~Shaposhnikov,}
\author[a,b]{S.S.~Vybin,}
\author[a]{A.F.~Bokhanov,}
\author[a]{M.Yu.~Kazakov,}
\author[a]{S.P.~Shlepnev,}
\author[a]{K.F.~Burdonov,}
\author[a]{A.A.~Soloviev,}
\author[a]{and M.V.~Starodubtsev}

\affiliation[a]{Federal Research Center Institute of Applied Physics, Russian Academy of Sciences\\
46 Ulyanova st., 603950, Nizhny Novgorod, Russia}
\affiliation[b]{National Research Lobachevsky State University of Nizhny Novgorod\\
23 Prospekt Gagarina, 603950, Nizhny Novgorod, Russia}

\emailAdd{ivizot@ipfran.ru}

\abstract{In this paper we report on a high current density ion beam profile diagnostics with a slit-based system as a reliable method, capable of high thermal load applications. The task arose in frames of a point-like neutron source development for neutron radiography. In previous research, it was suggested to construct such a system as a D-D neutron generator based on the high current gasdynamic ion source, which utilises the plasma of electron cyclotron resonance discharge sustained by powerful millimeter wave gyrotron radiation. This device is able to produce focused D$^+$ beams with a characteristic diameter of 1 mm, total current above 100 mA, and current density at a level of several A/cm$^2$. Study of such intense beams profile to obtain the best focusing efficiency and minimize neutron producing area appeared to be a challenging task. The paper also demonstrates the possibility of fast neutron imaging with a point-like powerful neutron generator (neutron yield on the level of 10$^{10}$~s$^{-1}$).}


\begin{document}
\maketitle
\flushbottom

\section{Introduction}
\label{sec:intro}

One of the most promising methods for modern non-invasive material diagnostics is a fast neutron radiography \cite{g1,g2}, which began to take off more than 80 years ago (in 1938) and is currently undergoing a period of tremendous growth. On the one hand, this is due to a wide spectrum of applications of its methods, in particular, it enables the study of composition and structure of objects containing both light and heavy elements, the possibility to detect structural defects (cavities, cracks, various inclusions with their dimensions ranging from several micrometers to tens of centimetres, monitoring of the dynamics (kinetics) of various physical, chemical, and technological processes under conditions when other types of monitoring cannot be used (e.g., in nuclear reactors, high-pressure and high-temperature chambers), monitoring of radioactive products and details, first and foremost, fuel elements of nuclear reactors, etc. On the other hand, the growing interest in neutron imaging is connected with the development of special visualization systems (e.g., imaging plates, scintillators) and detectors equipped with processing software systems and computer-controlled automatic neutron diffractometers. It is important to note that neutron radiography requires dedicated sources of neutrons. Either the sources delivering paraxial neutron beams with low velocity angular spread or point-like neutron sources are only suitable for the imaging purposes. One should mention an appreciable progress in the design of neutron sources. At present, the main source of the directed neutron beam with an intensity required for applications is a highly collimated powerful neutron flux from large-scale accelerators like ESS \cite{v6}, IFMIF \cite{v7}, SNS \cite{v8} or nuclear reactors. In such systems, the total neutron flux reaches the level of 10$^{15}$ s$^{-1}$cm$^{-2}$, and the neutron flux density -- 10$^8$-10$^9$ s$^{-1}\cdot$cm$^{-2}$ at the exit of a collimator with a characteristic aperture of 10 cm \cite{v3,v4,v5}.

Other types of neutron sources, e.g. conventional deuterium-deuterium (D-D) and deuterium-tritium (D-T) neutron generators \cite{v9,v10}, sources based on linear accelerators and RFQs \cite{v11,v12}, a point-like neutron sources based on powerful femtosecond lasers \cite{v13,v14} are significantly inferior to large scale facilities in terms of the neutron yield. Compact devices could be beneficial in case of point-like neutron emission, making it possible to avoid huge particle losses in a collimator. Relatively cheap and simple D-D and D-T neutron generators with a size of the emitting region of 2 mm \cite{v15} provide neutron fluxes at the level of 10$^7$ s$^{-1}$ (D-D) and 10$^9$ s$^{-1}$ (D-T). A noticeable neutron yield of up to 10$^8$ neutrons per pulse or 10$^7$ neutrons per 1 joule of the input laser energy was achieved in experiments with laser-plasma source with a characteristic size of 1 mm \cite{v16,v17,v18,v19}. In these experiments, 0.1 J pulses at a repetition rate of 1 kHz were used, yielding the time-averaged neutron flux on the level of 10$^7$ s$^{-1}$.

The possibility of creating a powerful point-like source of fast neutrons in a D-D (or D-T) generator scheme has been discussed in recent works \cite{g8,g9,g10}. It was proposed to use a sharp focusing of a high-current deuterium ion beam with an energy of up to 100 keV for bombardment of a deuterium (or tritium) loaded target. Then, 2.45 MeV (or 14.1 MeV) neutron flux is proportional to the ion beam current, whereas the size of the emitting region is solely determined by the ion beam cross-section, i.e. its focusing efficiency. The latter depends on the quality of the impinging deuterium beam, i.e. its emittance. This proposal arose from the development \cite{g15,g16,g17,g18} of high-current ion sources based on a discharge sustained by powerful electromagnetic radiation of millimeter wavelength gyrotrons under conditions of electron cyclotron resonance in open magnetic traps. The use of a high-power millimeter radiation allows to increase the discharge plasma density up to 10$^{13}$ - 10$^{14}$ cm$^{-3}$), which is by more than an order of magnitude higher when compared to the conventional ECR ion sources. At such plasma density level, the transition to a quasi-gasdynamic plasma confinement regime occurs, characterized with a short plasma lifetime (on the level of several tens of microseconds), which in turn makes it possible to form high current high-quality deuterium beams with a record brightness (deuterium beams with 450 mA current, 800 mA/cm$^2$ current density, normalized RMS emittance 0.07 $\pi\cdot$mm$\cdot$mrad, and a molecular ion fraction of less than 10 \% were obtained \cite{g15,g16,g17,g18,g19,g20}). Experimental studies reported in refs.~\cite{g22,g23}, confirmed the possibility of a point-like neutron source creation based on the principle described above, demonstrating a neutron yield of 10$^{10}$ s$^{-1}$ from the emitting region of 1 mm.

\begin{figure}[htbp]
	\centering 
	\includegraphics[width=0.8\textwidth]{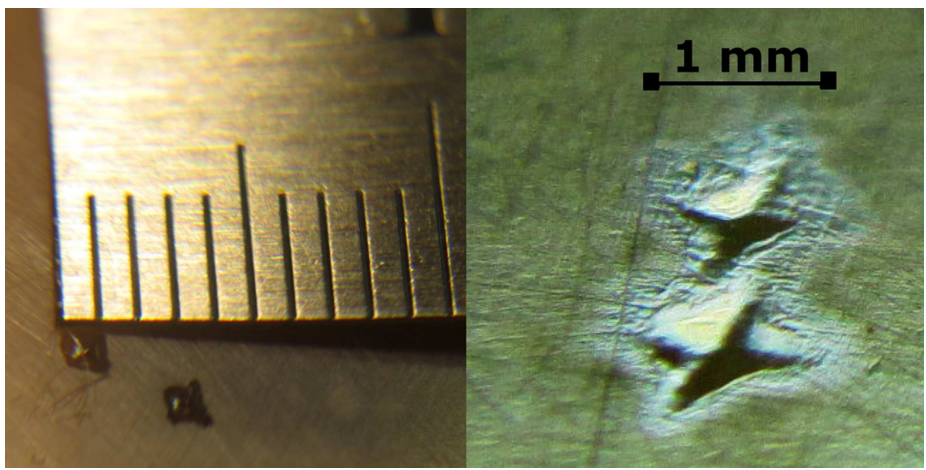}
	\caption{\label{fig:1} Scintillator defects formed after a single pulse.}
\end{figure}

The core element of the described point-like neutron generator is the formation system of converging high-current D$^+$ ion beam, since the focusing system determines the size of the emitting area of the generator and, therefore, the spatial resolution of neutronography diagnostic methods. The development of optimal focusing system requires adequate methods of measuring the spatial distribution of ion beam intensity. The latter was measured in [11, 15] by means of recording the scintillator luminescence, placed in a focal plane of the focusing system, assuming that the luminescence intensity is proportional to the ion beam current density. This method yields only qualitative estimation of the beam parameters and does not allow to analyze the focusing efficiency thoroughly. The main downside of the scintillator technique is that it is incapable of handling ion flux densities used in the experiments. Cavities formation was observed during the experiment with characteristic size of less than 1 mm on the scintillator surface (see Figure \ref{fig:1}).

The depth of the cavities is approximately equal to their width, which allows us to assume that this method may be corrupting the beam profile measurement results (note that for the ion flux densities used in the experiments, a specific energy deposition exceeded 10$^6$ J/cm$^3$ per pulse; given that the heat conductivity of the scintillator is very low, this leads to evaporation of the material and, consecutively, noticeable modification of the crystal surface). Thus, the problem of determining the optimal conditions for formation of converging high-current beams with extreme possible compression requires the development of more reliable methods of the beam spatial structure measurement. We propose a system of movable slit diaphragms made of aluminum, equipped with the Faraday cup for ion current measurement. The objective of this work is to develop such measurement method for high-current beams and determine the optimal parameters for the best focusing of a D$^+$ ion beam by a magnetic lens.


\section{Experimental setup}
\label{sec:exp}

\begin{figure}[htbp]
	\centering 
	\includegraphics[width=1.0\textwidth]{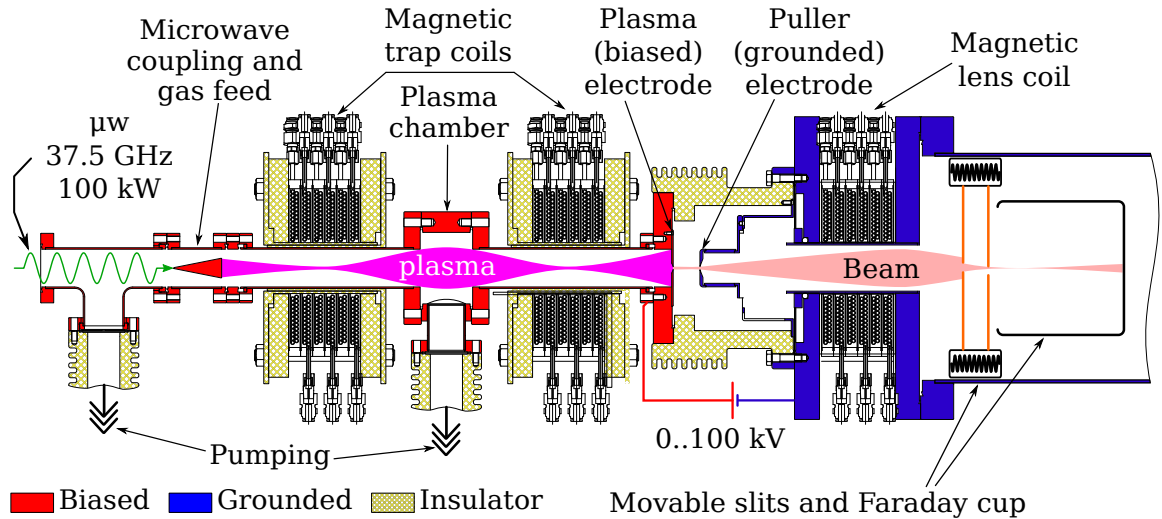}
	\caption{\label{fig:2} Experimental layout.}
\end{figure}

The experiments were conducted at SMIS-37 facility \cite{g23}. The schematic setup layout is shown in Figure~\ref{fig:2}. A pulsed 37.5 GHz gyrotron radiation with power of 100 kW was used for the plasma heating. Microwave pulse duration was equal to 1.5 ms at a repetition rate of 0.1 Hz. Microwave radiation was injected into the vacuum volume through the quartz window and focused afterwards to the center of a plasma chamber 250 mm long with 38 mm vacuum bore. The plasma chamber was enclosed with two coils placed 15 cm apart, providing an open plasma magnetic trap with a mirror configuration. The duration of coils current pulse was equal to 11 ms, inducing the peak field of 4 T at magnetic mirrors. The mirror ratio of the trap (B$_{max}$/B$_{min}$) was equal to~5. The plasma forming gas (deuterium) was supplied by the pulsed solenoid valve through the gas feed line embedded into the microwave coupling system, providing axial gas injection. All systems were synchronized so that the discharge was ignited at the peak of the magnetic field pulse.

The formation of the deuterium ion beam was realized using the electrostatic diode extraction system, consisting of two flat electrodes: plasma electrode (biased) and puller electrode (grounded) (see Fig.~\ref{fig:1}). In described experiments, the plasma of the quasi-gasdynamic discharge had the density of $\sim$2$\cdot$10$^{13}$ cm$^{-3}$, which together with electron temperature $\sim$50 eV led to the ion current density of several (up to 10) amperes per cm$^2$ in the magnetic mirror. The electric field on the order of 1~MV/cm would have been needed to form the ion beam, which is not technically feasible.
Despite the plasma being highly collisional, its flux follows the magnetic field in the region of interest \cite{shap}, allowing for easy way to tune the flux density in the extraction region by changing its position relative to the magnetic field. Thus, the plasma electrode was placed 100 mm downstream the beam line, where the trap magnetic field and, therefore, the plasma flux density decreased by a factor of 7 down to 1 A/cm$^2$, which falls within the voltage range available at the facility (up to 100 kV).
It is convenient to describe the extraction system with 3 numbers: A1-D-A2, where A1 is the aperture of the plasma electrode, A2 is the aperture of the puller electrode, and D is the interelectrode gap, all dimensions are given in mm. In the experiments two extraction systems were used: 5–10–7 and 5–5–5.
The total beam current was measured with a Faraday cup equipped with a shielding electrode for secondary electrons damping with a potential of -100 V with respect to the collector.
Two Spellman ST100P12 power supplies connected in parallel were used as a source of the high voltage biasing the extractor (a maximum voltage of 100 kV and a current of up to 240 mA). With the parameters of the discharge and extraction system optimized, such a source allowed generating ion beams with a total current of up to 200 mA. The aperture of the lens was equal to 76 mm, the lens length was 100 mm, and the maximum field intensity was 3 T. The transverse distribution of the high-current ion beam was measured by the double-slit scanner, developed for this purpose, in the focal area of the magnetic lens. The scanner consists of two orthogonal movable slits having a width of 500 $\mu$m, that are installed in the vacuum chamber perpendicularly to the beam axis and are equipped with remotely controlled step motors and feedback gauges.
The collector (the same Faraday cup used to measure the total beam current) that detects the ion current having passed through the aperture formed by the orthogonal slits is installed behind them.
Additionally, the first slit surface current was recorded, being proportional to the total current of the beam and allowing for the beam tuning (except for the regime of sharp focusing, when significant portion of the beam has passed through the slits aperture). High-precision mechanics (worm-gear drives, limit switches, etc.) and step motors manufactured by NEC were used to make the movable pieces of the assembly. A combination of used step motors and worm-gear drives ensured a slit-shifting accuracy of 150 $\mu$m.

\begin{figure}[htbp]
	\centering 
	\includegraphics[width=0.8\textwidth]{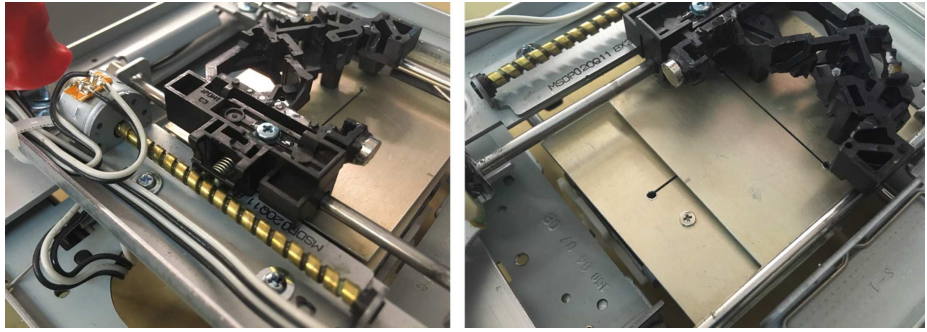}
	\caption{\label{fig:3} The system of movable plates with slits for the ion beam diagnostics.}
\end{figure}

The system is controlled remotely by means of a specially developed controller and the software responsible for setting, controlling, and calibrating the slit positions in the experiment. During the experiment, the ion beam was focused on the front slit of the scanner placed at a distance of 100 mm from the magnetic lens. The optimal focusing was achieved at the magnetic field of the lens being equal to 1 T. Figure \ref{fig:3} shows photos of the system elements.

\subsection{Ion beam distribution measurements}
The transverse distributions of the ion beam current in the focal region were studied following the gradient descent procedure. First, the maximum of the beam current was found with respect to one of the coordinates. Then, the maximum with respect to the other coordinate was found, after which the process was repeated for the first coordinate. After the precise coordinates of the maximum had been found, the intensities were determined along two orthogonal lines, which crossed at the beam intensity maximum, found earlier. Figure 4 shows characteristic transverse distributions of the ion current, which were obtained by using an extractor with 5-mm apertures in the plasma electrode and 7-mm apertures in the puller, and an interelectrode distance of 10 mm (5-10-7) for an extraction of 40, 50 and 60 kV.

\begin{figure}[htbp]
	\centering 
	\includegraphics[width=0.8\textwidth]{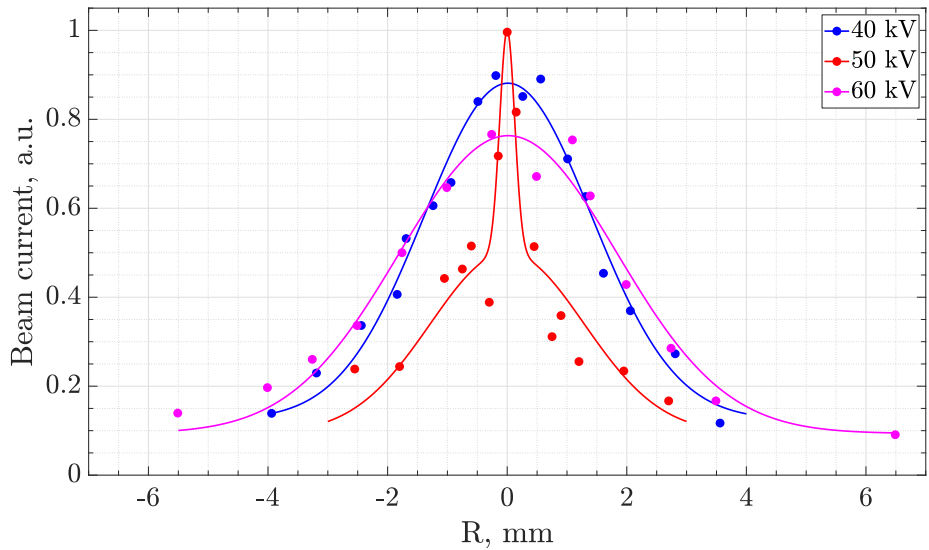}
	\caption{\label{fig:4} The transverse ion beam current density profiles at the extraction voltage 40 kV (blue), 50 kV (red) and 60 kV (green). Dots correspond to the experimental data, the solid curves correspond to the bi-Gaussian fit, covering both the D$^+$ ion and admixtures.}
\end{figure}

\begin{figure}[htbp]
	\centering 
	\includegraphics[width=0.8\textwidth]{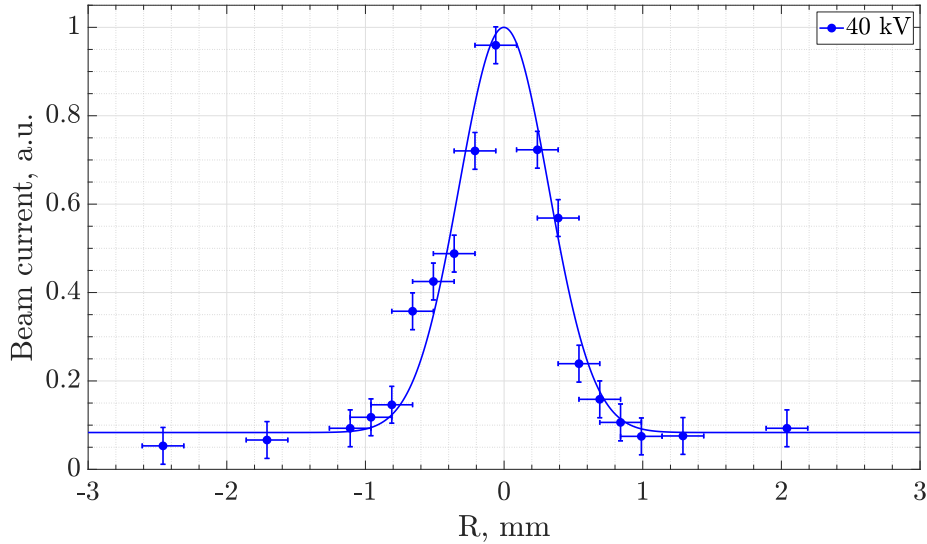}
	\caption{\label{fig:5} The transverse ion beam current density profile obtained at the extraction voltage of 40 kV and the optimized extraction system of 5 mm plasma electrode, 5 mm puller electrode and 5 mm interelectrode gap.}
\end{figure}

An optimal value of the focusing magnetic field was chosen for each voltage. It is seen in Figure~\ref{fig:4} that for the used configuration of the extraction system, the minimum transverse size of the ion beam is reached at an extraction voltage of 50 kV. The presence of the extraction voltage optimum is explained by the fixed longitudinal position of the slit scanner along the ion beam propagation path, since the focal plane of the magnetic lens also shifts along the longitudinal coordinate, as the beam energy changes. The position of the scanner, which was used in the experiment, corresponds to the position of the focal plane for a beam energy of 50 keV. Figure~\ref{fig:5} presents the distribution of ion beam with the minimum transverse size, which was obtained by using the extraction system (with a 5-mm aperture in the plasma electrode, a 5-mm aperture in the puller, and an interelectrode gap of 5 mm) for a beam energy of 40 keV.

The minimum size of the beam (full width at half maximum) is 0.8 mm. Apparently, the slit method has higher spatial resolution when compared to the scintillator-based method. Moreover, the spatial resolution of the measurement can be improved by using thinner slits. Note that even using slit scanner, we were not able to measure and/or generate beams with their transverse sizes comparable with simulated results in the experiments. The numerical simulation of the ion beam focusing performed with the open-source IBSimu software \cite{ibsimu} and allowing for the space-charge compensation, show that when a high-field magnetic lens is used, beams with an experimentally achieved emittance may be compressed to a diameter of 200 $\mu$m. At the same time, the experimentally measured beam diameter is significantly larger. The difference between the experiments and simulations can be due to presence of admixtures in the real beam while simulations were done for the pure D$^+$ beam. Admixtures in the real beam, which may not only add the background to the observed transversal distribution, but also affect the plasma meniscus shape, which is detrimental for ion trajectories simulation.

\subsection{Fast neutron imaging}
\begin{figure}[htbp]
	\centering 
	\includegraphics[width=1.0\textwidth]{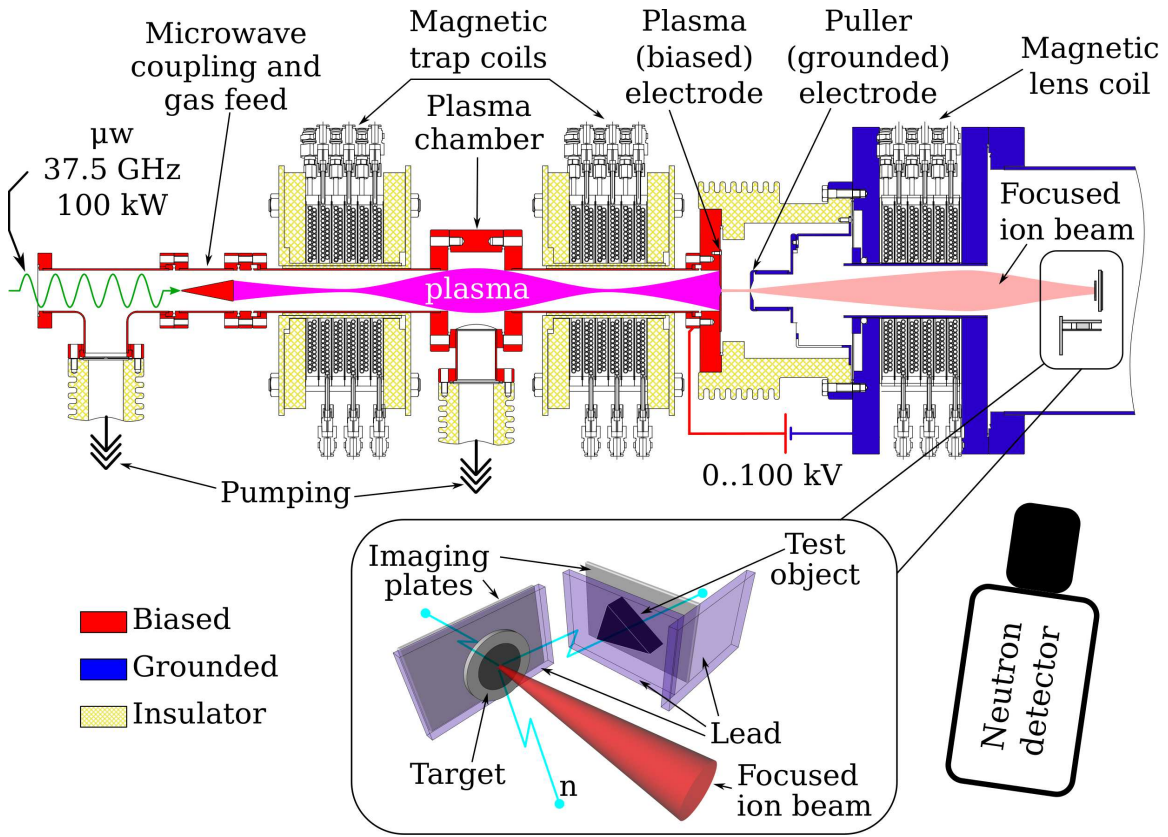}
	\caption{\label{fig:6} Scheme of the experiment on fast neutron imaging.}
\end{figure}

To test the possibility of obtaining images in fast neutron flux, the slit beam profile scanner was replaced with a neutron-generating target (see Figure~\ref{fig:6}). The target consisted of 10 $\mu$m of titanium deposited on a 1 mm thick molybdenum plate and saturated with D$_2$ to the level of TiD$_{1.2}$. The neutron flux intensity was measured using a calibrated MKS-AT1117M neutron detector based on a $^3$He proportional counter enclosed within a HDPE neutron moderator. The detector was located at a distance of 1.5 m from the neutron source at a considerable distance from the concrete walls of the room to mitigate the influence of neutron back-scattering. A 3 mm thick lead enclosure was used to shield the detector from plasma bremsstrahlung. Neutron flux measurements confirmed its growth with both ion beam energy and current and reached the value of 1.2$\cdot$10$^{10}$ s$^{-1}$ at the beam current of 50 mA and the beam energy of 75 keV (the total number of neutrons per pulse was $\sim$ 1.8$\cdot$10$^7$ on average). Assuming the isotropic neutron emission (which seems reasonable, as the beam energy is 75 keV only) from the area of  $\pi\cdot$0.1$^2$ cm$^2$, the neutron flux density at the target was equal to 3.8$\cdot$10$^{11}$ s$^{-1}\cdot$cm$^{-2}$.

The imaging plates by Fuji Computed Radiography, normally used in medical and scientific x-ray imaging, were utilized to visualize fast neutron flux coming from a point-like neutron source. The operating principle of imaging plates \cite{g2} is based on the effect of photo-stimulated luminescence. First, the imaging plates were exposed to the high-energy radiation (namely, neutrons in our experiments). Excited electrons, originating from the proton recoil chain, remained trapped in the color centers of the lattice. Imaging plates used in the experiment were made of the 200~$\mu$m phosphor layer covering a polyester film substrate. The layer consisted of barium fluorobromide, containing trace amounts of bivalent europium as color centers. Next, the plates were transferred into a scanner, where they were illuminated with an 630 nm laser, which forced the trapped electrons to the conducting band. Photo-induced luminescence at a wavelength of 400 nm was then recorded by a CCD matrix of the scanner.

Experiments on neutron imaging were performed with the relatively low ion beam current of 30 mA at 75 keV to ensure the target temperature would stay below the critical level of 300 \textdegree C. 500~pulses were accumulated, providing the total exposure time of $\sim$0.5~s. The first imaging plate was placed directly behind a neutron-generating target with a 2 mm thick lead plate in between, which prevented the plate from being exposed to the bremsstrahlung, coming from the plasma and the electrodes being bombarded by energetic electrons. Experiments performed in the same geometry, but with hydrogen as the plasma forming gas, had confirmed the sufficient thickness of the lead. The total distance between the first imaging plate and the target was 5 mm (see Fig.~\ref{fig:6}).

A triangular prism made of high density polyethylene (HDPE) with sides of 21$\times$15$\times$21.5~mm and thickness of 10~mm was used as a test object.
The material of the test object is largely composed of hydrogen, therefore it interacts effectively with the neutron flux. The test object was placed aside at a distance of 25 mm from the target. The second imaging plate was placed behind the test object; the whole assembly was shielded with 2 mm thick lead.

\begin{figure}[htbp]
	\centering 
	\includegraphics[width=1.0\textwidth]{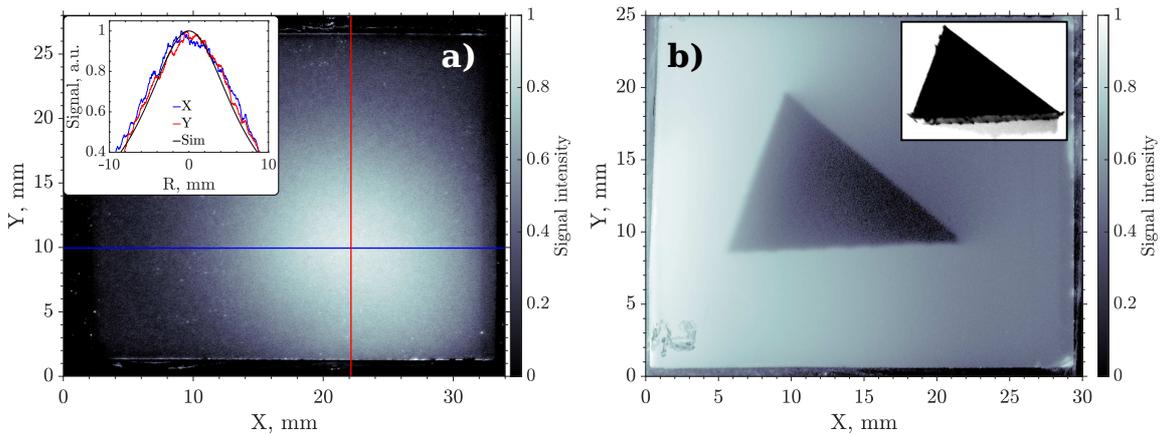}
	\caption{\label{fig:7} Pictures obtained with imaging plates.}
\end{figure}

The image of the neutron source itself is obtained at the first imaging plate is shown in Figure~\ref{fig:7}a. The center of the image, i.e. the maximum intensity point appeared to be shifted, apparently due to a misalignment of the assembly.
Linear intensity distributions along horizontal (blue curve) and vertical (red curve) dimensions are plotted in Figure~\ref{fig:7}a. A black line shows analytical distribution obtained within the simplest assumption that the image intensity is inversely proportional to the distance between the given image pixel and the source squared, and the source was assumed to be the circle 2 mm in diameter.
Such a simple approach yielded a good agreement with measurements, thus indirectly confirming the ion beam profile measurements.

The image obtained at the second plate is shown in Figure~\ref{fig:7}b. A picture of the test object (a photo of which is also shown in Figure~\ref{fig:7}b) is clearly seen. The image is positive, apparently due to effective neutron scattering within the test object material leading to weakening of the neutron flux and reducing its energy. It is of note that the level of image contrast appeared to be quite high, allowing to resolve fine details visible at the test object bottom edge.

\section{Conclusion}
A two-slits analyzer of spatial distribution of high-current ion beam intensity was developed, allowing to study the characteristics of a deuterium ion beam with a current of more than 50 mA and a characteristic diameter of 1 mm, analyze the distribution of its intensity over transverse coordinates with a spatial resolution of about 150 $\mu$m, and determine the optimal conditions (i.e., the values of the focusing magnetic field and the extracting electrostatic field, which are optimal for the specified configuration of the extracting electrodes and the position of the focal plane). Test experiments on the fast neutron imaging demonstrated the prospects of implementing fast neutron radiography using a point-like (characteristic size of the emitting region $\sim$1~mm) powerful (neutron yield at the level of 10$^{10}$ s$^{-1}$) D-D neutron generator, based on the ECR discharge with a quasi-gasdynamic plasma confinement mode, sustained in the mirror magnetic trap by powerful electromagnetic radiation of the millimeter wavelength range.

\acknowledgments
The work was supported by Russian Science Foundation, grant \#16-19-10501.

\bibliographystyle{JHEP}
\bibliography{ms}
\end{document}